\documentclass[a4paper, 10pt, conference]{ieeeconf}%
\usepackage{bm}
\usepackage{amssymb}
\usepackage{epsfig}
\usepackage{subfigure}
\usepackage{amsmath}
\usepackage{graphicx}
\usepackage{epstopdf}
\usepackage{amsfonts}
\usepackage{color}
\usepackage{multirow}%
\providecommand{\U}[1]{\protect\rule{.1in}{.1in}}
\newtheorem{problem}{\textbf{Problem}}

\IEEEoverridecommandlockouts
\begin{document}

\title{{\LARGE \textbf{Decentralized Optimal Control in Multi-lane Merging for
Connected and Automated Vehicles}}}
\author{Wei Xiao, Christos G. Cassandras and Calin Belta\thanks{Supported in part by
NSF under grants ECCS-1509084, DMS-1664644, CNS-1645681, IIS-1723995,
CPS-1446151, by AFOSR under grant FA9550-19-1-0158, by ARPA-E's NEXTCAR
program under grant DE-AR0000796 and by the MathWorks.}\thanks{The authors are
with the Division of Systems Engineering and Center for Information and
Systems Engineering, Boston University, Brookline, MA, 02446, USA
\texttt{{\small \{xiaowei, cgc, cbelta\}@bu.edu}}}}
\maketitle

\begin{abstract}
We address the problem of optimally controlling Connected and Automated
Vehicles (CAVs) arriving from two multi-lane roads and merging at multiple
points where the objective is to jointly minimize the travel time and energy
consumption of each CAV subject to speed-dependent safety constraints, as well
as speed and acceleration constraints. This problem was solved in prior work
for two single-lane roads. A direct extension to multi-lane roads is limited
by the computational complexity required to obtain an explicit optimal control
solution. Instead, we propose a general framework that converts a multi-lane
merging problem into a decentralized optimal control problem for each CAV in
a less-conservative way. To accomplish this, we employ a joint
optimal control and barrier function method to efficiently get an optimal
control for each CAV with guaranteed satisfaction of all constraints.
Simulation examples are included to compare the performance of the proposed
framework to a baseline provided by human-driven vehicles with results showing
significant improvements in both time and energy metrics.

\end{abstract}

\thispagestyle{empty} \pagestyle{empty}

%%%%%%%%%%%%%%%%%%%%%%%%%%%%%%%%%%%%%%%%%%%%%%%%%%%%%%%%%%%%%%%%%%%%%%%%%%%%%%%%

\section{INTRODUCTION}

Traffic management at merging points (usually, highway on-ramps) is one of the
most challenging problems within a transportation system in terms of safety,
congestion, and energy consumption, in addition to being a source of stress
for many drivers \cite{Schrank2015, Tideman2007, Waard2009}. Advances in
next-generation transportation system technologies and the emergence of
Connected and Automated Vehicles (CAVs) have the potential to drastically
improve a transportation network's performance by better assisting drivers in
making decisions, ultimately reducing energy consumption, air pollution,
congestion and accidents. An overview of automated vehicle-highway systems was provided in
\cite{Varaiya1993}.

Most research work just focuses on the single lane merging problem \cite{Mukai2017, Milanes2011, Domingues2018}, with limited work done in the multi-lane merging problem. In our recent work \cite{Wei2019ACC}, we addressed the merging problem through
a decentralized optimal control (OC) formulation and derived explicit
analytical solutions for each CAV when no constraints are active. We have
extended the solution to include constraints \cite{Xiao2018}, in which case
the computational cost depends on the number of constraints becoming active;
we have found this to become potentially prohibitive for a CAV to determine
through on-board resources. In addition, our analysis has thus far assumed no
noise in the vehicle dynamics and sensing measurements, and the dynamics have
precluded nonlinearities.

To address the limitations above, one can adopt on-line control methods such
as Model Predictive Control (MPC) (e.g., \cite{Cao2015,
Mukai2017,Ntousakis2016}) or the Control Barrier Function (CBF) method
\cite{Ames2019, Xiao2019}. MPC is very effective for problems with
simple (usually linear or linearized) dynamics, objectives and constraints.
Unlike MPC, the CBF method does not use states as decision variables in its
optimization process; instead, any continuously differentiable state
constraint is mapped onto a new constraint on the control input and can ensure
forward invariance of the associated set, i.e., a control input that satisfies
this new constraint is guaranteed to also satisfy the original constraint.
This allows the CBF method to be effective for complex objectives, nonlinear
dynamics, and constraints. We have adopted this approach to the single-lane
merging problem in recent work \cite{Wei2019} and shown that it provides good
approximations of the analytically obtained OC solutions. To account for both
optimality and computational complexity, we developed a \emph{joint optimal
control and barrier function} \emph{(OCBF)} controller in \cite{Wei2019itsc}
for a two-lane merging problem. The implementation of this approach is hard
for multi-lane merging, especially in determining the safety constraints that
a CAV has to satisfy. The common approach to avoid such complex safety
constraint determination is to treat an entire conflict area as a point (i.e.,
only allow one CAV to enter the conflict area when there are possible
collisions), which is conservative (e.g., for an intersection, see
\cite{Zhang2018}). Alternatively, the conflict area can be partitioned
according to lane intersections and a tree search approach may be used to find
a feasible path for each CAV \cite{Xu2019}; this approach is limited by the
computational complexity due to the high-dimensional search space involved.

The contribution of this paper is to show how we can transform a multi-lane
merging problem into a multi-point merging problem in a simpler and
less-conservative way. Specifically, we first determine the merging points
that a CAV must pass through and construct queueing tables maintained by a
coordinator associated with the merging area. Using a simple search through
these tables, we determine the safe merging and rear-end safety constraints
that a CAV has to satisfy, hence transforming the multi-lane merging problem
into a decentralized optimal control problem for each CAV. Finally, we use the
aforementioned OCBF method to solve these optimal control problems. The main
advantages of the proposed framework lie in the optimality it provides, its
computational efficiency, safety guarantees, and good generalization
properties for even more complex traffic scenarios. Simulation results of the
proposed framework have shown significantly better performance compared to
human-driven vehicles.

%%%%%%%%%%%%%%%%%%%%%%%%%%%%%%%%%%%%%%%%%%%%%%%%%%%%%%%%%%%%%%%%%%%%%%%%%%%%%%%%

\section{PROBLEM FORMULATION}

\label{sec:problem}

\begin{figure*}[ptbh]
\centering
\includegraphics[scale=0.36]{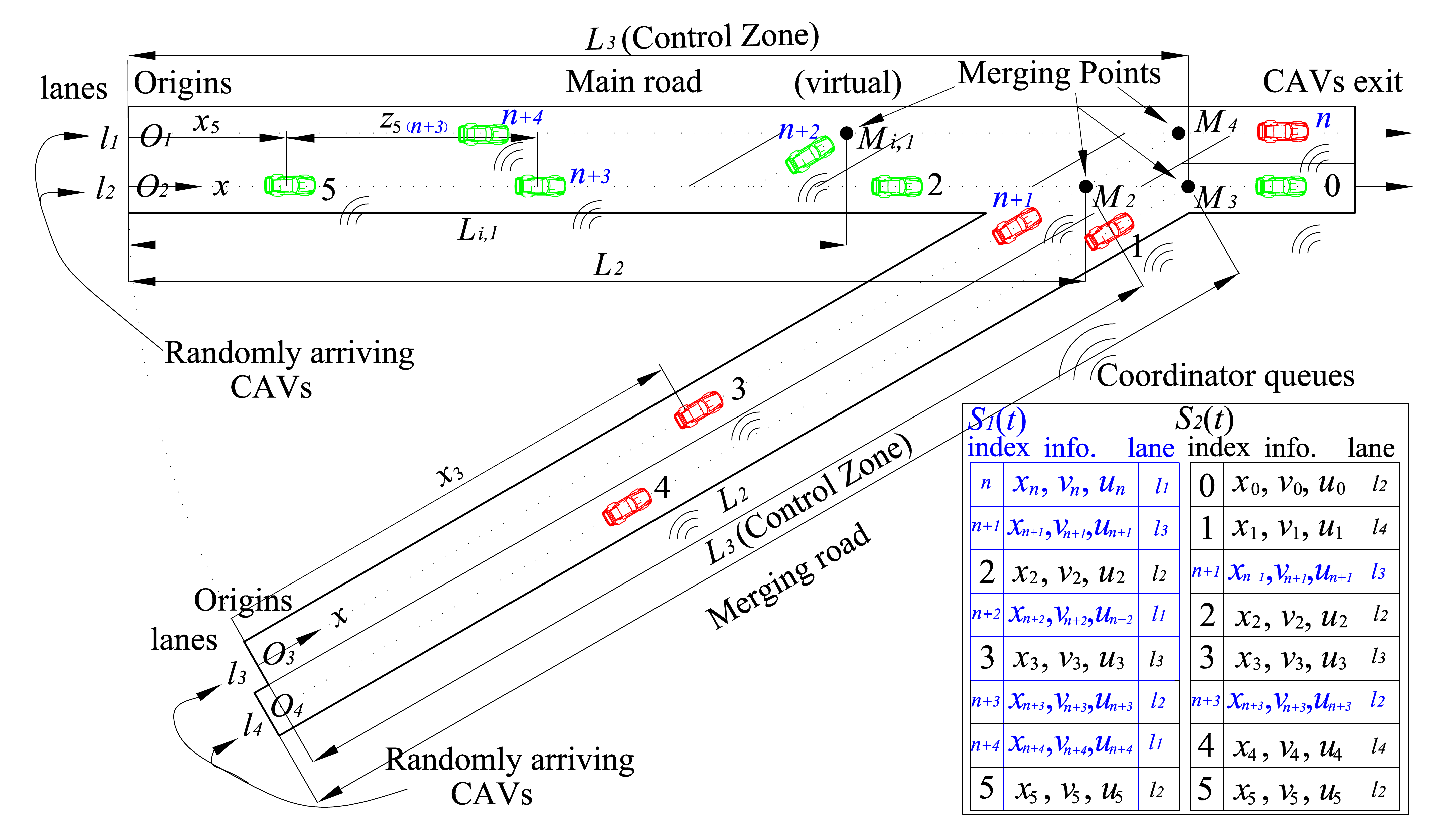} 
\vspace{-3mm}
\caption{The multi-lane merging problem.
Collisions may happen at the merging points $M_{i,1}, M_{2}, M_{3}, M_{4}$.}%
\label{fig:merging}%
\vspace{-5mm}
\end{figure*}

The multi-lane merging problem arises when traffic must be joined from two
different roads, usually associated with a main and a merging road as shown in
Fig.\ref{fig:merging}. Each road has two lanes (as we will see, the same
modeling method can be applied to more than two lanes). We label the lanes
$l_{1},l_{2}$ and $l_{3},l_{4}$ for the main and merging roads respectively,
with corresponding origins $O_{1},O_{2},O_{3},O_{4}$. Only the CAVs in lane
$l_{2}$ can change lanes to $l_{1}$. In addition, the CAVs in lane $l_{3}$
have the option to merge into either lane $l_{1}$ or $l_{2}$ (the main benefit
being that the CAV in $l_{3}$ can surpass a group of CAVs in $l_{4}$ when
$l_{4}$ is congested). Finally, the CAVs in lane $l_{4}$ can only merge to
$l_{2}$.

In our original single-lane merging problem \cite{Wei2019ACC} only lanes
$l_{2},l_{4}$ are involved and the only merging point is $M_{3}$ in
Fig.\ref{fig:merging}. Here, CAVs from lanes $l_{1},l_{2},l_{3},l_{4}$ may
merge at the three fixed merging points $M_{2},M_{3},M_{4}$. In addition, a
CAV from lane $l_{2}$ may merge into $l_{1}$ at an arbitrary merging point
$M_{i,1}$, as long as this point is located prior to $M_{2}$. We consider the
case where all traffic consists of CAVs randomly arriving at the four lanes
joined at the Merging Points (MPs) $M_{i,1},M_{2},M_{3},M_{4}$ where a
collision may occur. The road segment from $O_{2}$ or $O_{4}$ to the merging
point $M_{3}$ has a length $L_{3}$ and is called the \emph{Control Zone (CZ)}.
The segment from $O_{1}$ to $M_{i,1}$ for CAV $i$ has a length $L_{i,1}$
(which is variable and depends on $i$). The segment from $O_{2}$ or $O_{3}$ to
$M_{2}$ has a length $L_{2}$.

We assume that CAVs do not overtake each other in the CZ (unless so dictated
by the CAV's controller to be developed in the sequel), that $L_{i,1}<L_{2}$,
and that the merging point $M_{4}$ is within the CZ. Moreover, note that if
the controller determines that a CAV needs to change lanes from $l_{2}$ to
$l_{1}$, then it has to travel an additional distance; we assume that this
extra distance is a constant $l>0$. The same constant applies to CAVs in lane
$l_{3}$ which choose to merge into $l_{1}$ at $M_{4}$ (as opposed to merging
into $l_{2}$).

A coordinator (typically a Road Side Unit (RSU)) is associated with the MP
$M_{3}$ whose function is to maintain First-In-First-Out (FIFO) queues of all
CAVs regardless of lanes based on their arrival time at the CZ and to enable
real-time communication with the CAVs that are in the CZ as well as the last
one leaving the CZ (in particular, the coordinator does not make control
decisions; this is done in decentralized fashion on-board each CAV). The FIFO
assumption (so that CAVs cross the MP in their order of arrival) is made for
simplicity and often to ensure fairness; however, it can be relaxed through
dynamic resequencing schemes as described, for example, in \cite{Zhang2018},
\cite{Wei2020ACC}. Since we have two lanes in the main road, we need two
queues to manage each CAV sequence leaving the CZ via $l_{1}$ and $l_{2}$
respectively, as shown in Fig. \ref{fig:merging}. Note that the number of
queues equals the number of lanes in the main road, thus this framework can be
easily extended to other multi-lane road traffic configurations, such as intersections.

Let $S_{1}(t),S_{2}(t)$ be the sets of the FIFO-ordered CAV indices associated
with the two possible CZ exit lanes $l_{1}$ and $l_{2}$. To maintain a single
unique index for each CAV, let $n>0$ be a large enough integer representing
the road capacity over $L_{3}$ in terms of the number of CAVs that can be
accommodated. Then, let the set of possible CAV indices in $S_{2}(t)$ be
$\{0,1,\ldots,n-1\}$ and that in $S_{1}(t)$ be $\{n,n+1,\ldots,2n-1\}$. Thus,
CAV $n+j$ $(j\in\mathbb{N)}$ belongs to $S_{1}(t)$. The CAVs indexed by $n$ or
$0$ are the ones that have just left the CZ from $l_{1},l_{2}$ respectively.
Let $N_{1}(t),N_{2}(t)$ be the cardinalities of $S_{1}(t),S_{2}(t)$,
respectively. Observe that the CAVs in any one queue may have a physical
conflict (i.e., collisions may happen) with the CAVs in the other queue only
in lanes $l_{2},l_{3}$, but not in lanes $l_{1},l_{4}$. Thus, we assign a
newly arriving CAV according to the following cases:

$(i)$ If a CAV arrives at time $t$ at lane $l_{1}$, it is assigned to
$S_{1}(t)$ with an index $n+N_{1}(t)$.

$(ii)$ If a CAV arrives at time $t$ at lane $l_{2}$, a decision is made (as
decsribed later) on whether it exits the CZ through $l_{2}$ or switches to
$l_{1}$ at $L_{i,1}$. This CAV is assigned to both $S_{1}(t)$ and $S_{2}(t)$
with the index $N_{2}(t)$ if it chooses to stay in $l_{2}$ (e.g., CAV $2$ in
Fig. \ref{fig:merging}) or the index $n+N_{1}(t)$ if it switches to $l_{1}$
(e.g., CAV $n+3$ in Fig. \ref{fig:merging}).

$(iii)$ If a CAV arrives at time $t$ at lane $l_{3}$, it is assigned to both
$S_{1}(t)$ and $S_{2}(t)$ with the index $n+N_{1}(t)$ if the control decision
is to merge to lane $l_{1}$ or the index $N_{2}(t)$ if it merges to lane
$l_{2}$.

$(iv)$ If a CAV arrives at time $t$ at lane $l_{4}$, it is assigned to
$S_{2}(t)$ with the index $N_{2}(t)$.

Note that in the above case $(ii)$, the index of the CAV arriving at $l_{2}$
is dropped from $S_{2}(t)$ (or $S_{1}(t)$) after it changes its lane to
$l_{1}$ at $M_{i,1}$ (or passes $M_{2}$). In the above case $(iii)$, the index
of the CAV arriving at lane $l_{3}$ is dropped from $S_{1}(t)$ (or $S_{2}(t)$)
after it passes $M_{2}$ if it chooses to merge into $l_{2}$ (or $l_{1}$). In
summary, the index of any CAV arriving at $O_{2}$ or $O_{3}$ will be dropped
from queue $S_{1}(t)$ or $S_{2}(t)$ after it passes its first MP. This is to
ensure a correct queue management corresponding to the fact that a CAV is
added to \emph{both} queues in the above cases $(ii)$ and $(iii)$. All CAV
indices in $S_{2}(t)$ decrease by one when a CAV passes MP $M_{3}$ and the CAV
whose index becomes $-1$ is dropped (similarly for $S_{1}(t)$, {the CAV
leaving the CZ through $M_{4}$ whose index becomes $n-1$ is dropped}). Observe
that this scheme allows any CAV $i\in S_{1}(t)$ to look up only queue table
$S_{1}(t)$ (similarly for $S_{2}(t)$ if $i\in S_{2}(t)$) in order to identify
all possible collisions with other CAVs, without any need to consider the
other queue.

The vehicle dynamics for each CAV $i\in S_{1}(t)\cup S_{2}(t)$ along the lane
to which it belongs takes the form
\begin{equation}
\left[
\begin{array}
[c]{c}%
\dot{x}_{i}(t)\\
\dot{v}_{i}(t)
\end{array}
\right]  =\left[
\begin{array}
[c]{c}%
v_{i}(t)+w_{i,1}(t)\\
u_{i}(t)+w_{i,2}(t)
\end{array}
\right]  , \label{VehicleDynamics}%
\end{equation}
where $x_{i}(t)$ denotes the distance to the origin $O_{1}$ or $O_{2}%
,O_{3},O_{4}$ along the lane that $i$ is located in when it enters the CZ,
$v_{i}(t)$ denotes the velocity, and $u_{i}(t)$ denotes the control input
(acceleration). Moreover, $w_{i,1}(t),w_{i,2}(t)$ denote two random processes
defined in an appropriate probability space to capture possible noise. We
consider two objectives for each CAV subject to three constraints, as detailed next.

\textbf{Objective 1} (Minimize travel time): Let $t_{i}^{0}$ and $t_{i}^{m}$
denote the time that CAV $i\in S_{1}(t)\cup S_{2}(t)$ arrives at the origin
$O_{1}$ or $O_{2},O_{3},O_{4}$ and the time that CAV $i$ leaves the CZ
(through either $M_{3}$ or $M_{4}$), respectively. We wish to minimize the
travel time $t_{i}^{m}-t_{i}^{0}$ for CAV $i$.

\textbf{Objective 2} (Minimize energy consumption): We also wish to minimize
the energy consumption for each CAV $i\in S_{1}(t)\cup S_{2}(t)$ expressed as
\begin{small}
\begin{equation}
J_{i}(u_{i}(t))=\int_{t_{i}^{0}}^{t_{i}^{m}}\mathcal{C}(u_{i}(t))dt,
\label{eqn:obj}%
\end{equation}
\end{small}where $\mathcal{C}(\cdot)$ is a strictly increasing function of its argument.

\textbf{Constraint 1} (Safety constraint): Let $i_{p}$ denote the index of
the CAV which physically immediately precedes $i\in S_{1}(t)\cup S_{2}(t)$ in the CZ (if one is
present). We require that the distance $z_{i,i_{p}}(t)\equiv x_{i_{p}%
}(t)-x_{i}(t)$ be constrained by:
\begin{equation}
z_{i,i_{p}}(t)\geq\varphi v_{i}(t)+\delta,\text{ \ }\forall t\in\lbrack
t_{i}^{0},t_{i}^{m}], \label{Safety}%
\end{equation}
where $\varphi$ denotes the reaction time (as a rule, $\varphi=1.8$ is used,
e.g., \cite{Vogel2003}). If we define $z_{i,i_{p}}$ to be the distance from
the center of CAV $i$ to the center of CAV $i_{p}$, then $\delta$ is a
constant determined by the length of these two CAVs (generally dependent on
$i$ and $i_{p}$ but taken to be a constant over all CAVs for simplicity).

\textbf{Constraint 2} (Safe merging): Let $t_{i}^{m_{p}},$ $p\in\{1,2,3,4\}$
denote the arrival time of CAV $i\in S_{1}(t)\cup S_{2}(t)$ (note that CAV $i$
will only pass at most two of these MPs) at the merging points $M_{i,1}%
,M_{2},M_{3},M_{4}$, respectively. There should be enough safe space at these
MPs for a merging CAV $i$ to cut in, i.e.,
\begin{small}
\begin{equation}
\label{SafeMerging}\begin{aligned} z_{i,j}(t_{i}^{m_{1}})&\geq\varphi v_{i}(t_{i}^{m_{1}})+\delta,\text{ \ }i\in S_{1}(t),\\ z_{i,j}(t_{i}^{m_{2}})&\geq\varphi v_{i}(t_{i}^{m_{2}})+\delta,\text{ \ }i\in S_{1}(t)\cup S_{2}(t),\\ z_{i,j}(t_{i}^{m_{3}})&\geq\varphi v_{i}(t_{i}^{m_{3}})+\delta,\text{ \ }i\in S_{2}(t),\\ z_{i,j}(t_{i}^{m_{4}})&\geq\varphi v_{i}(t_{i}^{m_{4}})+\delta,\text{ \ }i\in S_{1}(t), \end{aligned}
\end{equation}
\end{small}where $j\in S_{1}(t)\cup S_{2}(t)$ is the CAV that may collide with $i$ ($j$
may not exist) at the merging points $M_{i,1},M_{2},M_{3},M_{4}$. Observe that
since a CAV crosses at most two of the four MPs, CAV $i$ only needs to satisfy
the safe merging constraints above corresponding to the MPs that it will
actually cross (e.g., CAV $1$ in Fig. \ref{fig:merging} only needs to satisfy
the third constraint in (\ref{SafeMerging})). The index $j$ corresponding to
each $i$ is generally hard to determine; we will resolve this issue in the
next section through a conflict-point-based method.

\textbf{Constraint 3} (Vehicle limitations): Finally, there are constraints on
the speed and control for each $i\in S_{1}(t)\cup S_{2}(t)$:
\begin{equation}
\begin{aligned} v_{min} &\leq v_i(t)\leq v_{max}, \forall t\in[t_i^0,t_i^{m}],\\ u_{i,min}&\leq u_i(t)\leq u_{i,max}, \forall t\in[t_i^0,t_i^{m}], \end{aligned} \label{VehicleConstraints}%
\end{equation}
where $v_{max}>0$ and $v_{min}\geq0$ denote the maximum and minimum speed
allowed in the CZ, while $u_{i,min}<0$ and $u_{i,max}>0$ denote the minimum
and maximum control for each CAV $i$, respectively.

A common way to minimize energy consumption is by minimizing the control input
effort $u_{i}^{2}(t)$. By normalizing travel time and $u_{i}^{2}(t)$, and
using $\alpha\in\lbrack0,1)$, we construct a convex combination as follows:
\begin{small}
\begin{equation}
\begin{aligned}\min_{u_{i}(t)} J_i(u_i(t))= \int_{t_i^0}^{t_i^{m}}\left(\alpha + \frac{(1-\alpha)\frac{1}{2}u_i^2(t)}{\frac{1}{2}\max \{u_{max}^2, u_{min}^2\}}\right)dt \end{aligned}.
\label{eqn:energyobja}%
\end{equation}
\end{small}If $\alpha=1$, then we solve (\ref{eqn:energyobja}) as a minimum time problem.
Otherwise, by defining $\beta\equiv\frac{\alpha\max\{u_{\max}^{2},u_{\min}%
^{2}\}}{2(1-\alpha)}$ and multiplying (\ref{eqn:energyobja}) by the constant
$\frac{\beta}{\alpha}$, we have:
\begin{small}
\begin{equation}
\min_{u_{i}(t)}J_{i}(u_{i}(t))=\beta(t_{i}^{m}-t_{i}^{0})+\int_{t_{i}^{0}%
}^{t_{i}^{m}}\frac{1}{2}u_{i}^{2}(t)dt, \label{eqn:energyobj}%
\end{equation}
\end{small}where $\beta\geq0$ is a weight factor that can be adjusted through $\alpha
\in\lbrack0,1)$ to penalize travel time relative to the energy cost. Then, we
have the following problem formulation:

\begin{problem}
\label{prob:merg} For each CAV $i\in S_{1}(t)\cup S_{2}(t)$ governed by
dynamics (\ref{VehicleDynamics}), determine a control law such that
(\ref{eqn:energyobj}) is minimized subject to (\ref{VehicleDynamics}),
(\ref{Safety}), (\ref{SafeMerging}), (\ref{VehicleConstraints}), given $t_{i}^{0}$ and the initial and final conditions $x_{i}(t_{i}%
^{0})=0$, $v_{i}(t_{i}^{0})$, $x_{i}(t_{i}^{m})$.
\end{problem}

\section{Multi-lane Merging Problem Solution}

\label{sec:ocbf}

We now show how to decompose Problem \ref{prob:merg} into a multi-point
merging problem for each CAV and use the CBF method to account for constraints
while tracking a CAV trajectory obtained through OC. We also take advantage of
the robustness to noise that the CBF approach offers.

However, determining the exact merging constraints in (\ref{SafeMerging}) that
a CAV $i\in S_{1}(t)\cup S_{2}(t)$ has to satisfy is challenging since there
are four lanes and the traffic is asymmetric. This is even harder for more
lanes and other scenarios, such as intersections. Using the approach
introduced in \cite{Wei2019ACC} and considering the multi-lane merging problem
in Fig. \ref{fig:merging}, there are 15 cases, making this hard to implement.
Moreover, this approach does not scale well for more complicated cases.
Therefore, we propose a conflict-point based approach to simplify this
process, as described next.

\subsection{Lane Merging Determination Strategy}

\label{sec:lane} When a new CAV $i\in S_{1}(t)\cup S_{2}(t)$ arrives at
$O_{2}$ or $O_{3}$, it has the option of exiting the CZ through lane $l_{1}$
or $l_{2}$. In addition, if it arives at $O_{2}$ and decides to merge to
$l_{1}$, it must also determine the location of the variable MP $M_{i,1}$.

Let us begin with the first issue. Determining the lane from which a CAV should exit the CZ
may be addressed using the optimal dynamic resequencing method from
\cite{Wei2020ACC}, the only difference being that CAV $i$ has a binary
decision to make. Thus, we can solve a constrained OC problem as in
\cite{Wei2020ACC} (accounting for the possibility that one or more of the
speed, control and safety constraints becomes active) under each option. This
becomes computationally intensive; for example in the single-lane merging
problem we have found this to require 3 to 30sec in MATLAB \cite{Wei2020ACC},
and this will generally increase in the multi-lane merging problem at hand.
Although this remains an option (by seeking more effcient implemenation
algorithms to solve the underlying OC problem), in this paper we focus on
computational efficiency by adopting the following lane-merging decision
strategy: we seek to balance the expected number of CAVs in the two lanes in
order to improve the cost (\ref{eqn:energyobj}) on average. In a
queueing-theoretic context, this implies adopting a shortest-queue-first
policy which is known to be often optimal in terms of minimizing average
travel times. Thus, for any arriving CAV $i$ at $O_{2}$ or $O_{3}$ at
$t_{i}^{0}$:
\begin{equation}
i\in\left\{
\begin{array}
[c]{rcl}%
S_{1}(t), & \mbox{if }N_{1}(t_{i}^{0})<N_{2}(t_{i}^{0}) & \\
S_{2}(t), & \mbox{otherwise} &
\end{array}
,\text{ \ }t\in\lbrack t_{i}^{0},t_{i}^{m}].\right.  \label{eqn:lane}%
\end{equation}

Next, we address the issue of selecting the location of the MP $M_{i,1}$ for a
CAV $i$ arriving at $O_{2}$, if its decision is $i\in S_{1}(t)$ above. There
are three important observations to make: $(i)$ The \emph{unconstrained}
optimal control for such $i$ is independent of the location of $M_{i,1}$ since
we have assumed that lane-changing will only induce a fixed extra length $l$.
$(ii)$ The OC solution under the first safe-merging constraint in
(\ref{SafeMerging}) is better (i.e., lower cost in (\ref{eqn:energyobj})) than
one which includes an active rear-end safety constrained arc in its optimal
trajectory. This is because the former applies only to a single time instant
$t_{i}^{m_{1}}$ whereas the latter requires the constraint (\ref{Safety}) to
be satisfied over all $t\in\lbrack t_{i}^{0},t_{i}^{m_{1}}]$. It follows that
the merging point $M_{i,1}$ should be as close as possible to $M_{2}$ (i.e.,
$L_{i,1}$ should be as large as possible), since the safe-merging constraint
between $i$ and $i-1$ will become a rear-end safety constraint after $M_{i,1}%
$. $(iii)$ In addition, CAV $i$ arriving at $O_{2}$ may also be constrained by
its physically preceding CAV $i_{p}$ (if one exists) in lane $l_{2}$. In this
case, CAV $i$ needs to consider both the rear-end safety constraint with
$i_{p}$ and the safe-merging constraint with $i-1$. Thus, the solution is more
constrained (hence, more sub-optimal) if $i$ stays in lane $l_{2}$ after the
rear-end safety constraint due to $i_{p}$ becomes active. We conclude that in
this case CAV $i$ should merge to lane $l_{1}$ when the rear-end safety
constraint with $i_{p}$ in lane $l_{2}$ first becomes active, i.e., $L_{i,1}$
is determined by
\begin{equation}
L_{i,1}=x_{i}^{\ast}(t_{i}^{a}) \label{Li1}%
\end{equation}
where $x_{i}^{\ast}(t)$ denotes the unconstrained optimal trajectory of CAV
$i$ (as determined in Sec. \ref{sec:OCBF}), and $t_{i}^{a}\geq t_{i}^{0}$ is
the time instant when the rear-end safety constraint first becomes active
between $i$ and $i_{p}$ in lane $l_{2}$; if this constraint never becomes
active, then $t_{i}^{a}=t_{i}^{m_{2}}$. The value of $t_{i}^{a}$ is determined
from (\ref{Safety}) by
\begin{equation}
x_{i_{p}}^{\ast}(t_{i}^{a})-x_{i}^{\ast}(t_{i}^{a})=\varphi v_{i}^{\ast}%
(t_{i}^{a})+\delta, \label{tia}%
\end{equation}
where $x_{i_{p}}^{\ast}(t),v_{i}^{\ast}(t)$ are the \emph{unconstrained}
optimal trajectory and optimal speed respectively of CAV $i_{p}$. If, however,
CAV $i_{p}$'s optimal trajectory includes a constrained arc, then (\ref{tia})
is only an approximation (in fact, an upper bound) of $t_{i}^{a}$. In summary,
if CAV $i$ never encounters a point on $l_{2}$ where its rear-end safety
constraint becomes active, we set $L_{i,1}=L_{2}$, otherwise $L_{i,1}$ is
determined through (\ref{Li1})-(\ref{tia}).

\subsection{Merging Constraint Determination Strategy}

\label{sec:merg_stra} 
\begin{figure}[ptbh]
	\vspace{-5mm}
\centering
\includegraphics[scale=0.2]{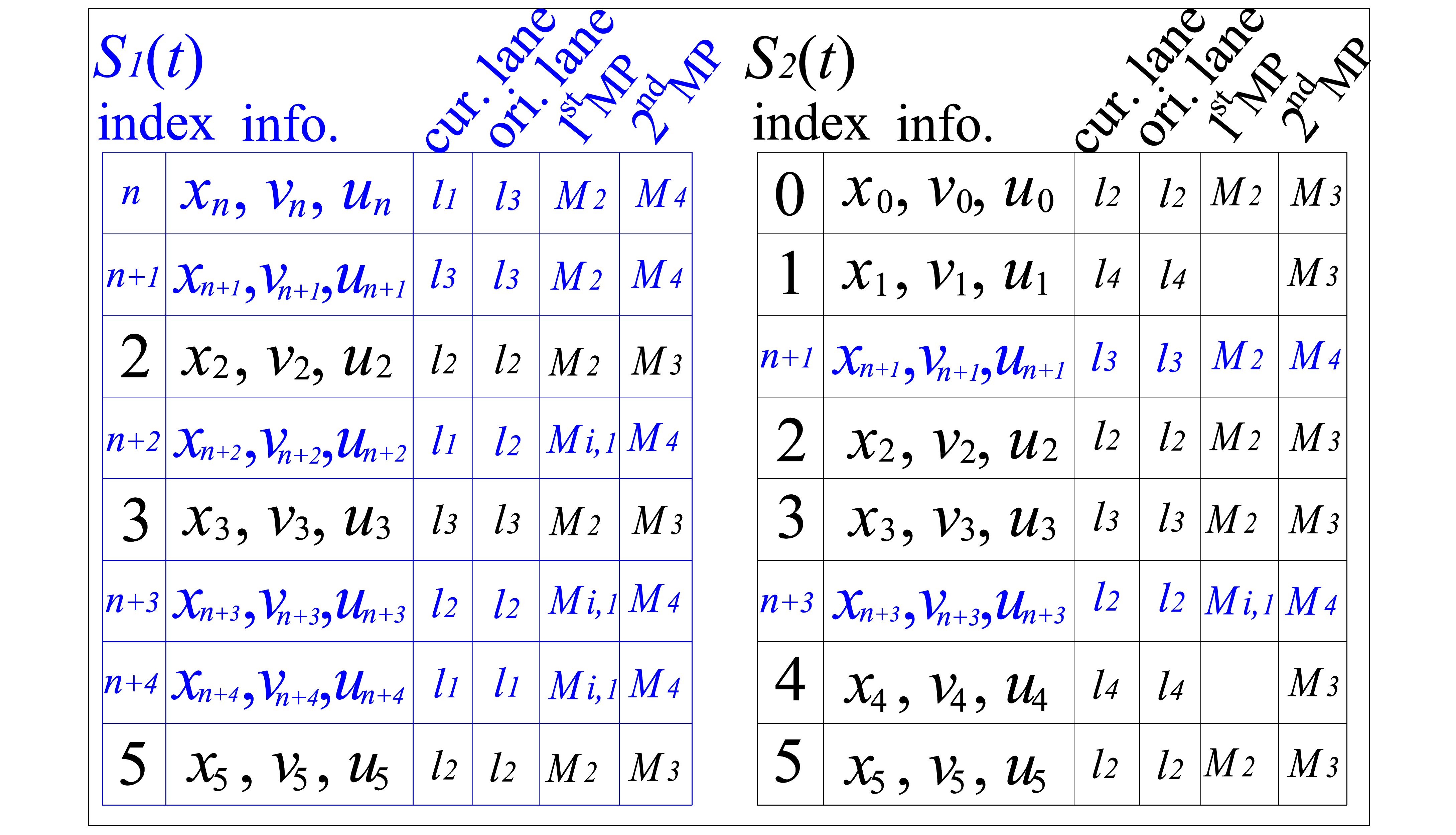}
\vspace{-5mm}
 \caption{The extended coordinator queue
tables.}%
\label{fig:queue}%
\vspace{-3mm}
\end{figure}

The CAVs arriving at lanes $l_{2},l_{3}$ will pass two
MPs. On the other hand, CAVs arriving at lane $l_{1}$ will pass either one or
two MPs (depending on whether $i$ and $i-1$ are in the same lane or not),
whereas all CAVs arriving at $l_{4}$ will pass only MP $M_{3}$. Moreover, CAVs
arriving at lanes $l_{2},l_{3}$ may pass through different MPs, depending on
which lane they choose to merge into following the strategy presented in the
last subsection. Since all MPs that a CAV has to pass are now determined, we
augment the FIFO queues in Fig. \ref{fig:merging} with the original lane and
the MP information for each CAV as shown in Fig. \ref{fig:queue}. The current
and original lanes are shown in the third and fourth column, respectively. The
last two columns indicate the first and second MPs for each CAV (note that all
CAVs arriving at lane $l_{4}$ and some CAVs arriving at lane $l_{1}$ have only
one MP, in which case the first MP is left blank). 

When a new CAV $i$ arrives at $O_{1}$ (or $O_{2},O_{3},O_{4}$) and has
determined whether it will merge into another lane or not (based on the last
subsection), it looks up the extended queue tables in Fig. \ref{fig:queue}
which already contain all prior CAV state and MP information. If $i\in
S_{1}(t)$, it looks up the extended FIFO queue $S_{1}(t)$, otherwise, it looks
up $S_{2}(t)$. From the \textit{current lane} column in Fig. \ref{fig:queue},
CAV $i$ can determine its current physically immediately preceding CAV $i_{p}$
if one exists. Moreover, CAV $i$ can determine the safe-merging constraints
that it should satisfy (i.e., with respect to which CAV $j$ in
(\ref{SafeMerging}) in the queue) upon its arrival at any origin.

The precise process through which each arriving CAV $i$ looks up each queue
$S_{1}(t)$ and $S_{2}(t)$ in Fig. \ref{fig:queue} is a follows. CAV $i$
compares its \textit{original lane} and MP information to that of every CAV in
each queue \emph{starting with the last row and moving up}. Depending on which
column (among the last three columns) matches first, there are four possible
cases (a much smaller number than 15 if the approach in \cite{Wei2019ACC},
\cite{Wei2019}, \cite{Wei2019itsc} were followed). \textit{This process
terminates the first time that any one of these four cases is satisfied at
some row.} If that does not happen, this implies that CAV $i$ does not have to
satisfy any safe-merging constraint. Let $type(i)\in\{1,2\}$ be such that
$type(i)=1$ if $i\geq n$ and $type(i)=2$ otherwise. Then, the four cases are:

$(1)$ All last three columns match first.

$(2)$ [$1^{st}$ MP column matches with $j\in S_{1}(t)$ (or $S_{2}(t)$) first]
\& [$type(i)=type(j)$].

$(3)$ [$1^{st}$ MP column matches with $j\in S_{1}(t)$ (or $S_{2}(t)$) first)
\& [$type(i)\neq type(j)$].

$(4)$ The $2^{nd}$ MP column matches first.

When a new CAV $i$ arrives and $i\in S_{1}(t)$ (similarly if $i\in S_{2}(t)$),
it first checks for case $(1)$. If case $(1)$ is satisfied, this means that
CAV $i_{p}\in S_{1}(t)$ is the physically immediately preceding CAV all the
way through the CZ. Thus, CAV $i$ only has to satisfy the safety constraint
(\ref{Safety}) with respect to $i_{p}$, i.e., it just follows CAV $i_{p}$. For
example, $i=n+3,$ $i_{p}=n+2$ in Fig. \ref{fig:merging}.

If case $(2)$ is first satisfied for CAV $i\in S_{1}(t)$ (or $S_{2}(t)$), then
CAV $i$ has to satisfy the first or the second safe-merging constraint in
(\ref{SafeMerging}) with CAV $j\in S_{1}(t)$. Moreover, it has to satisfy the
safety constraint (\ref{Safety}) with $i_{p}\in S_{1}(t)$, where $i_{p}$ is
found by the first matched row in the \emph{current lane} column of Fig.
\ref{fig:queue}. Since $type(i)=type(j)$, the first or the second safe merging
constraint in (\ref{SafeMerging}) will become the safety constraint
(\ref{Safety}) after CAV $i$ passes the first MP, therefore, there is no
further safe-merging constraint at the second MP $M_{3}$ or $M_{4}$ (CAV $i$
just follows CAV $j$ after the first MP). For example, $i=n+4,$ $j=n+3$ in
Fig. \ref{fig:merging}.

In case $(3)$, CAV $i\in S_{1}(t)$ (or $S_{2}(t)$) has to satisfy the first or
the second safe-merging constraint in (\ref{SafeMerging}) with CAV $j\in
S_{1}(t)$. Moreover, it has to satisfy the safety constraint (\ref{Safety})
with $i_{p}\in S_{1}(t)$, where $i_{p}$ is found by the first matched row in
the \emph{current lane} column of Fig. \ref{fig:queue}. Since $type(i)\neq
type(j)$, CAV $i$ cannot follow CAV $j$ after the first MP since $i$ and $j$
will merge into different lanes. Therefore, CAV $i$ also has to satisfy the
safe-merging constraint with CAV $k\in S_{1}(t)$ (where $k$ is found by the
first matched row in the $2^{nd}$ MP column of Fig. \ref{fig:queue}). For
example, $i=2,$ $j=n+1,$ $k=1$ in Fig. \ref{fig:merging}. Observe that it is
possible that $i_{p}=k$, in which case the third safe-merging constraint in
(\ref{SafeMerging}) is a redundant constraint.

As for the last case, CAV $i\in S_{1}(t)$ (or $S_{2}(t)$) has to satisfy the
first or the second safe-merging constraint in (\ref{SafeMerging}) with CAV
$j\in S_{1}(t)$. In addition, it has to satisfy the third or the fourth
safe-merging constraint in (\ref{SafeMerging}) with CAV $k$, determined by the
first matched row in the $1^{st}$ MP column of Fig. \ref{fig:queue}), and it
has to satisfy the safety constraint (\ref{Safety}) with $i_{p}\in S_{1}(t)$,
where $i_{p}$ is found by the first matched row in the \emph{current lane}
column of Fig. \ref{fig:queue}). For example, $i=5,$ $j=4,$ $k=3$ (and
$i_{p}=n+3$ at the current time, but note that this will change to $i_{p}=2$
after CAV $n+3$ merges into lane $l_{1}$) in Fig. \ref{fig:merging}.

If none of the four cases above is satisfied, then CAV $i$ does not have to
satisfy any safe-merging constraint. In summary, a newly arriving
CAV may have to satisfy at most three safety (or safe-merging) constraints in
Fig. \ref{fig:merging}. If the corresponding $k$ or $i_{p}$ is not found in
the above cases, then the related safe-merging or safety constraint is skipped.

\textbf{Updating }$S_{1}(t)$\textbf{ and }$S_{2}(t)$. Observe that while the
MP information in the last two columns of each queue in Fig. \ref{fig:queue}
remains unchanged, the same is not true for the \emph{current lane}
information. More precisely, the two queues need to be updated whenever one of
the following four events takes place: $(i)$ A new CAV arrives at the CZ and
is added to one or both queues. $(ii)$ {A CAV $i\in S_{2}(t)$ (or $S_{1}(t)$)
leaves the CZ causing the index of any CAV $j\in S_{1}(t)\cup S_{2}(t)$ with
$type(j)=2$ (or $type(j)=1$) to decrease by 1 and the CAV whose index is $-1$
(or $n-1$ in $S_{1}(t)$) is removed from $S_{2}(t)$ (or $S_{1}(t)$). Note that
CAV $-1$ only appears in $S_{2}(t)$ (CAV $n-1$ only appears in $S_{1}(t)$), as
discussed in Sec. \ref{sec:problem}.} $(iii)$ A CAV changes lanes, causing an
update in the \textit{current lane} column in Fig. \ref{fig:queue}. This event
is important because the value of $i_{p}$ for any CAV $i$ already in a queue
may change, since its original $i_{p}$ may merge into another lane. $(iv)$ A
CAV overtake event when a CAV passes $M_{3}$ or $M_{4}$. This may occur when a
CAV $i\in S_{2}(t)$ (or $S_{1}(t)$) overtakes $i-1\in S_{1}(t)\cup S_{2}(t)$
when the two CAVs pass different MPs without conflict. Thus, if $i$ passes
$M_{3}$ or $M_{4}$ and $i-1$ is still in one of the queues, we need to
re-order $S_{2}(t)$ (or $S_{1}(t)$) according to the incremental position
order, so that CAV $i+1$ can properly identify its $(i+1)_{p}$. For example,
consider $i=4,$ $i-1=n+3,$ $i+1=5$ in queue $S_{2}(t)$ of Fig.
\ref{fig:merging}. CAV 4 can overtake $n+3$, and its current lane will become
$l_{2}$ when it passes $M_{3}$. When this happens, CAV 5 may mistake CAV 4 as
its $i_{p}$ by looking at the new current lane entry for it, which is now in
$l_{2}$. In reality, $i_{p}=n+3$ as long as CAV $n+3$ is still in lane $l_{2}%
$. This is avoided by re-ordering queue $S_{2}(t)$ according to the position
information when this event occurs (i.e., swapping rows for CAVs $4$ and $n+3$).

We can now solve Problem \ref{prob:merg} for all $i\in S_{1}(t)\cup S_{2}(t)$
in a decentralized way, in the sense that CAV $i$ can solve it using only its
own local information (position, velocity and acceleration) along with that of
its \textquotedblleft neighbor\textquotedblright\ CAVs found through the above
four cases. This is described next.

\subsection{Joint Optimal and Barrier Function Controller}

\label{sec:OCBF}Once a newly arriving CAV $i\in S_{1}(t)\cup S_{2}(t)$ has
determined all the safe merging constraints it has to satisfy as described in
the last subsection, it can solve problem (\ref{eqn:energyobj}) subject to
these constraints along with the rear-end safety constraint (\ref{Safety}) and
the state limitations (\ref{VehicleConstraints}). Obtaining a solution to this
constrained optimal control problem is computationally intensive in the
single-lane merging problem \cite{Wei2019ACC}, and is obviously more
computationally intensive in the multi-lane merging problem, since a CAV may
have to satisfy two safe-merging constraints. Therefore, we will employ the
joint optimal control and barrier function (OCBF) controller developed in
\cite{Wei2019itsc} to account for all constraints.

We begin by noting that the distances from $O_{2},O_{3},O_{4}$ to $M_{2}$ or
$M_{3}$ are all the same, while the distances from $O_{1},O_{2}$ to $M_{i,1}$
or $M_{4}$ (or from $O_{1},O_{3}$ to $M_{4}$) are different since the lane
change behavior will induce an extra $l$ distance (a CAV moving from $M_{2}$
to $M_{4}$ is equivalent to a lane change). Therefore, we need to perform a
coordinate transformation for those CAVs that are in different lanes (e.g.,
$l_{2}$ and $l_{1}$) and will merge into the same lane (e.g., $l_{1}$). In
other words, when $i\in S_{1}(t)$ obtains information for $j\in S_{1}(t)$ from
queue 1, the position information $x_{j}(t)$ is transformed by (using the
\emph{original lane} information in Fig. \ref{fig:queue}):
\begin{equation}
x_{j}(t):=\left\{
\begin{array}
[c]{rcl}%
x_{j}(t)+l, & \mbox{if [$i$ in $l_2$ or $l_3$] \& [$i-1$ in $l_1$],} & \\
x_{j}(t)-l, & \mbox{if [$i$ in $l_1$] \& [$i-1$ in $l_2$ or $l_3$]}, & \\
x_{j}(t), & \mbox{Otherwise}. &
\end{array}
\right.  \label{eqn:trans}%
\end{equation}
Note that the coordinate transformation (\ref{eqn:trans}) only applies to CAV
$i$ obtaining information on $j$ from $S_{1}(t)$, and does not apply to the
coordinator. Moreover, recall that after CAV $i\in S_{1}(t)$ merges into lane
$l_{1}$ from lane $l_{2}$ or $l_{3}$, it will be removed from $S_{2}(t)$.

Next, we briefly review the OCBF approach in \cite{Wei2019itsc} as it applies
to our problem. Problem (\ref{eqn:energyobj}) was solved in \cite{Wei2019ACC}
for the single-lane merging problem and no noise in (\ref{VehicleDynamics})
and the \emph{unconstrained} solution gives the following optimal control,
speed, and position trajectories:
\begin{equation}
u_{i}^{\ast}(t)=a_{i}t+b_{i}\label{Optimal_u}%
\vspace{-2mm}
\end{equation}%
\begin{equation}
v_{i}^{\ast}(t)=\frac{1}{2}a_{i}t^{2}+b_{i}t+c_{i}\label{Optimal_v}%
\vspace{-2mm}
\end{equation}%
\begin{equation}
x_{i}^{\ast}(t)=\frac{1}{6}a_{i}t^{3}+\frac{1}{2}b_{i}t^{2}+c_{i}%
t+d_{i}\label{Optimal_x}%
\end{equation}
where $a_{i}$, $b_{i}$, $c_{i}$ and $d_{i}$ are integration constants that can
be solved along with{ $t_{i}^{m}$} by the following five nonlinear algebraic
equations:
\begin{equation}
\begin{aligned} &\frac{1}{2}a_i\cdot(t_i^0)^2 + b_it_i^0 + c_i = v_i^0,\\ &\frac{1}{6}a_i\cdot(t_i^0)^3 + \frac{1}{2}b_i\cdot(t_i^0)^2 + c_it_i^0+d_i = 0,\\ &\frac{1}{6}a_i\cdot(t_i^{m})^3 + \frac{1}{2}b_i\cdot(t_i^{m})^2 + c_it_i^{m}+d_i = L_k,\\ &a_it_i^{m} + b_i = 0,\\ &\beta + \frac{1}{2}a_i^2\cdot(t_i^{m})^2 + a_ib_it_i^{m} + a_ic_i = 0. \end{aligned}\label{OptimalSolInA}%
\end{equation}
where the third equation is the terminal condition for the total distance
traveled $L_{k}$ on a lane given by $L_{k}=L_{3}+l$ if $i$ is in $l_{2}$ or
$l_{3}$ and chooses to merge into $l_{1}$; otherwise, $L_{k}=L_{3}$. This
solution is computationally very efficient to obtain (less than 1 sec in
MATLAB). We use this unconstrained OC solution as a \emph{reference} to be
tracked by a controller which uses CBFs to account for all the constraints
(\ref{Safety}), (\ref{VehicleConstraints}) and (\ref{SafeMerging}), hence this
combines an OC solution with CBFs and is referred to as an OCBF controller.
The only complication here is that the safe merging constraints in
(\ref{SafeMerging}) have to be converted to continuously differentiable forms
so as to be used in the CBF method. Thus, we use the same technique as in
\cite{Wei2019} to convert (\ref{SafeMerging}) into:
{\small
\begin{equation}
\begin{aligned} z_{i,j}(t)&\geq\Phi_{1}(x_{i}(t)) v_{i}(t)+\delta, i\in S_{1}(t), t\in [t_{i}^{0},t_{i}^{m_{1}}],\\ z_{i,j}(t)&\geq\Phi_{2}(x_{i}(t)) v_{i}(t)\!+\!\delta, i\!\in\! S_{1}(t)\!\cup\! S_{2}(t), t\!\in\![t_{i}^{0},t_{i}^{m_{2}}],\\ z_{i,j}(t)&\geq\Phi_{3}(x_{i}(t)) v_{i}(t)+\delta, i\in S_{2}(t), t\in [t_{i}^{0},t_{i}^{m_{3}}],\\ z_{i,j}(t)&\geq\Phi_{4}(x_{i}(t)) v_{i}(t)+\delta, i\in S_{1}(t), t\in [t_{i}^{0},t_{i}^{m_{4}}], \end{aligned}\label{SafeMergingc}%
\end{equation}}
where CAV $j$ is determined through the merging constraint determination
strategy of the last subsection and $\Phi_{p}:\mathbb{R}\rightarrow
\mathbb{R},$ $p\in\{1,2,3,4\}$ denote strictly increasing functions that
satisfy $\Phi_{p}(0)=-\frac{\delta}{v_{i}^{0}}$ (where $v_{i}^{0}$ denotes the
initial speed at the origin) and $\Phi_{p}(L_{p})=\varphi$ (for $p=1$, we set
$L_{1}=L_{i,1}$ since $L_{i,1}$ has been determined in Sec. \ref{sec:lane}).
Thus, we see that at $t=t_{i}^{m_{p}}$ when $x_{i}(t_{i}^{m_{p}})=L_{p}$ all
constraints in (\ref{SafeMergingc}) conform to the safe-merging constraints
(\ref{SafeMerging}), {and $z_{i,i_{p}}(t)=0$ at $t=t_{i}^{0}$ (all CAVs could
arrive at the same time at the four origins)}. Since the selection of
$\Phi_{p}(\cdot)$ is flexible, for simplicity, we define it to have the linear
form $\Phi_{p}(x_{i}(t))=(\varphi+\frac{\delta}{v_{i}^{0}})\frac{x_{i}%
(t)}{L_{p}}-\frac{\delta}{v_{i}^{0}}$.

The OCBF controller aims to track the OC solution (\ref{Optimal_u}%
)-(\ref{Optimal_x}) while satisfying all constraints (\ref{Safety}),
(\ref{VehicleConstraints}) and (\ref{SafeMergingc}). To accomplish this, first
let $\bm x_{i}(t)\equiv(x_{i}(t),v_{i}(t))$. Referring to the vehicle dynamics
(\ref{VehicleDynamics}), let $f(\bm x_{i}(t))=[x_{i}(t),0]^{T}$ and $g(\bm
x_{i}(t))=[0,1]^{T}$. Each of the seven constraints in (\ref{Safety}),
(\ref{VehicleConstraints}) and (\ref{SafeMergingc}) can be expressed as
$b_{k}(\bm x_{i}(t))\geq0$, $k\in\{1,\ldots,7\}$ where each $b_{k}(\bm
x_{i}(t))$ is a CBF. For example, we have $b_{1}(\bm x_{i}(t))=z_{i,i_{p}%
}(t)-\varphi v_{i}(t)-\delta$ for the rear-end safety constraint
(\ref{Safety}). In the CBF approach, each of the continously differentiable
\emph{state} constraints $b_{k}(\bm x_{i}(t))\geq0$ is mapped onto another
constraint on the \emph{control} input such that the satisfaction of this new
constraint implies the satisfaction of the original constraint $b_{k}(\bm
x_{i}(t))\geq0$. The forward invariance property of this method \cite{Ames2019, Xiao2019} ensures that a control input that satisfies the new
constraint is guaranteed to also satisfy the original constraint. In
particular, each of these new constraints takes the form
\begin{equation}
\begin{aligned} L_fb_k(\bm x_i(t)) + L_gb_k(\bm x_i(t))u_i(t) + \gamma( b_k(\bm x_i(t))) \geq 0, \end{aligned}\label{eqn:cbf}%
\end{equation}
where $L_{f},L_{g}$ denote the Lie derivatives of $b_{k}(\bm x_{i}(t))$
along $f$ and $g$ (defined above from the vehicle dynamics) respectively and
$\gamma(\cdot)$ denotes a class of $\mathcal{K}$ function
\cite{Khalil2002} (typically, linear and quadratic functions). As an
alternative, a Control Lyapunov Function (CLF) \cite{Ames2019} $V(\bm
x_{i}(t))$ can also be used to track (stabilize) the optimal speed trajectory
(\ref{Optimal_v}) through a CLF constraint of the form%
\begin{small}
\begin{equation}
\begin{aligned} L_fV(\bm x_i(t)) + L_gV(\bm x_i(t))u_i(t) + \epsilon V(\bm x_i(t)) \leq e_{i}(t), \end{aligned}\label{CLF}%
\end{equation}
\end{small}where $\epsilon>0$ and $e_{i}(t)$ is a relaxation variable that makes this
constraint soft. As is usually the case, we select $V(\bm x_{i}(t))=(v_{i}%
(t)-v_{ref}(t))^{2}$ where $v_{ref}(t)$ is the reference speed to be tracked
(specified below). Therefore, the OCBF controller solves the following
problem: 
{\small
\begin{equation}
\min_{u_{i}(t),e_{i}(t)}J_{i}(u_{i}(t),e_{i}(t))\!=\!\int_{t_{i}^{0}}%
^{t_{i}^{m}}\!\left(  \beta e_{i}^{2}(t)\!+\!\frac{1}{2}(u_{i}(t)\!-\!u_{ref}%
(t))^{2}\right)  dt,\label{eqn:ocbf}%
\end{equation}
}subject to the vehicle dynamics (\ref{VehicleDynamics}), the CBF constraints
(\ref{eqn:cbf}) and the CLF constraint (\ref{CLF}). The obvious selection for
speed and acceleration reference signals is $v_{ref}(t)=v_{i}^{\ast}(t)$,
$u_{ref}(t)=u_{i}^{\ast}(t)$, but we
select{\small
\begin{equation}
v_{ref}(t)=\frac{x_{i}^{\ast}(t)}{x_{i}(t)}v_{i}^{\ast}(t)\label{SpeedTrack}%
\end{equation}%
\begin{equation}
u_{ref}(t)=\frac{x_{i}^{\ast}(t)}{x_{i}(t)}u_{i}^{\ast}(t)\label{ControlTrack}%
\end{equation}}so as to provide position feedback to automatically reduce (or eliminate) the
tracking position error, since the optimal solutions in (\ref{Optimal_u}%
)-(\ref{Optimal_x}) depend on the position (alternative forms of $v_{ref}(t)$,
$u_{ref}(t)$ are possible as shown in \cite{Wei2019itsc}).

We refer to the resulting control $u_{i}(t)$ in (\ref{eqn:ocbf}) as the OCBF
control. The solution to (\ref{eqn:ocbf}) is obtained {by discretizing the
time interval }$[t_{i}^{0},t_{i}^{m}]$ with time steps of length $\Delta$ and
solving {(\ref{eqn:ocbf}) over }$[t_{i}^{0}+k\Delta,t_{i}^{0}+(k+1)\Delta]$,
$k=0,1,\ldots$, {with $u_{i}(t),e_{i}(t)$ as decision variables held constant
over each such interval. Consequently, each such problem is a Quadratic
Problem (QP) since we have a quadratic cost and a number of linear constraints
on the decision variables at the beginning of each time interval. The solution
of each such problem gives an optimal control $u_{i}^{\ast}(t_{i}^{0}%
+k\Delta)$}, $k=0,1,\ldots$, allowing us{ to update (\ref{VehicleDynamics}) in
the }$k^{th}${ time interval. This process is repeated until CAV $i$ leaves
the CZ.} The OCBF control can also deal with constraint violation due to noise
in the dynamics included in {(\ref{VehicleDynamics})}, as shown in
\cite{Wei2019itsc}.

\begin{figure}[ptbh]
	\vspace{-3mm}
\centering
\includegraphics[scale=0.10]{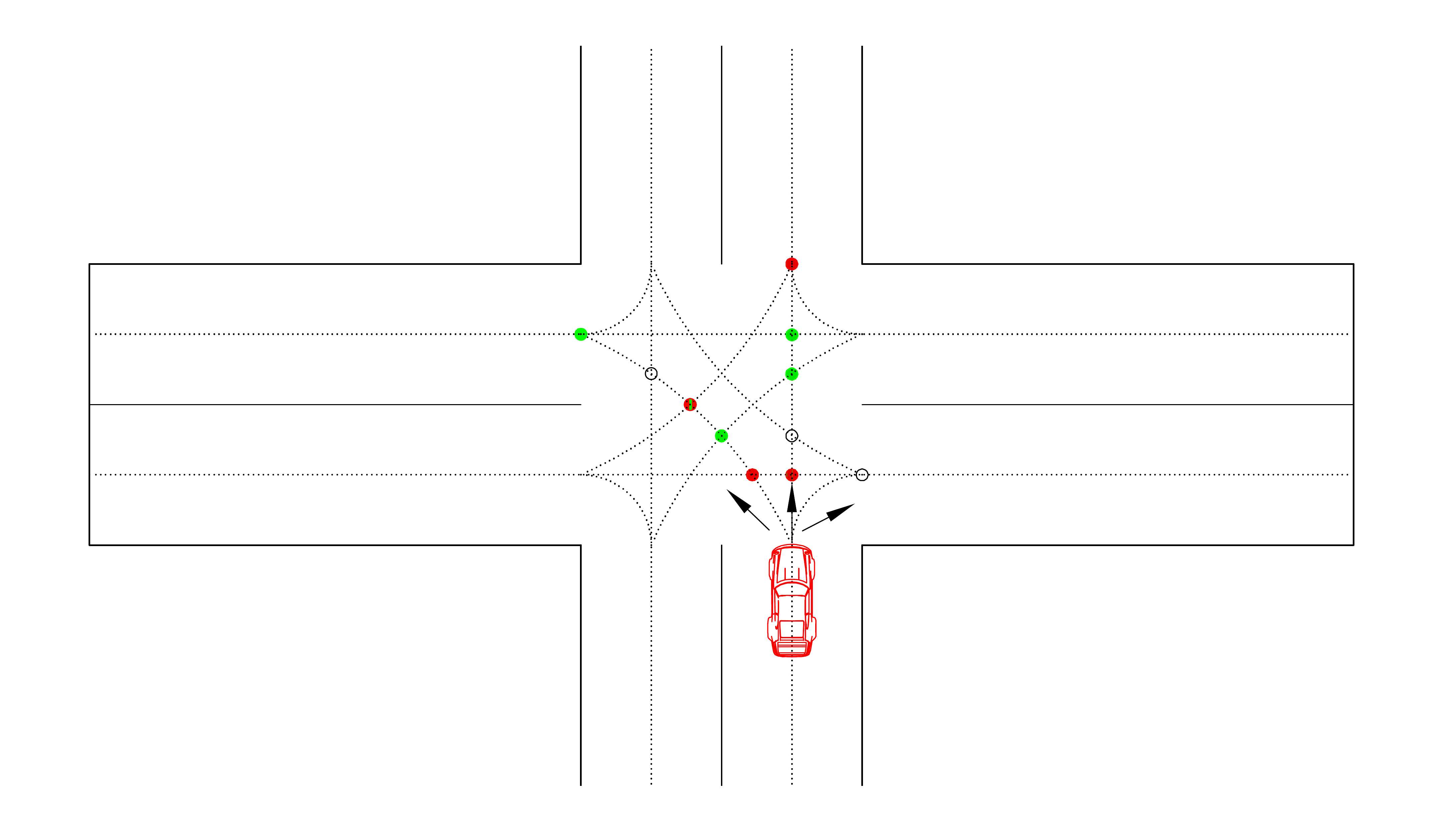} 
\vspace{-3mm}
\caption{The intersection
scenario.}%
\label{fig:inter}%
\vspace{-3mm}
\end{figure}

\textbf{Remark (Framework Generalization).} We can generalize the framework of
any traffic scenario that involves multiple lanes leading to conflict zones
beyond the merging configuration of Fig. \ref{fig:merging}. Suppose there are
$N_{e}\in\mathbb{N}$ exit lanes and at most $n\in\mathbb{N}$ merging
(conflict) points that a CAV may pass. Then, we can build $N_{e}$ FIFO queues
for any such scenario, with any new arriving CAV assigned to the queues whose
CAVs may have physical conflict with this new CAV. Then, according to the path
that this CAV will choose, we can identify all the merging points that it may
pass, and extend the FIFO queues with the proper order of passing merging
points similar to Fig. \ref{fig:queue}. Note that the same merging point may
appear at different columns in other rows (i.e., for other CAVs), so that the
matching approach proposed in Sec. \ref{sec:merg_stra} should compare all
other columns instead of just the same column as in the scenario of Fig.
\ref{fig:merging}. The number of possible cases $S_{n}$ (excluding the case
where $i_{p}$ is in the same lane as $i$ allthrough the CZ, as the case $(1)$
in Sec. \ref{sec:merg_stra}) is determined by
$
S_{n}=1+2S_{n-1}+S_{n-2}+S_{n-3}+\dots+S_{1},
$
where $S_{1}=1$. The number of all possible cases with respect to the number
of merging (conflict) points $n$ that a CAV may pass is given by $S_{n}+1$.
For example, a CAV in the intersection scenario shown in Fig. \ref{fig:inter}
may pass five merging points ($n=5$) (i.e., go straight or turn left) and
there are four exit lanes $N_{e}=4$. We have four FIFO queues, and extend them
in the form of Table \ref{fig:queue} by five MP columns. The number of
possible cases is 56, but a CAV can easily find all the safe merging
constraints (at most 5) that it needs to satisfy by looking up the extended
queue similar to the form in Table \ref{fig:queue}.

\section{SIMULATION RESULTS}

\label{sec:simulation}

All controllers in this section have been implemented using \textsc{MATLAB}
and we have used the Vissim microscopic multi-model traffic flow simulation
tool as a baseline for the purpose of making comparisons between our
controllers and human-driven vehicles adopting standard car-following models
used in Vissim. We used \textsc{quadprog} for solving QPs of the form
(\ref{eqn:ocbf}) and \textsc{ode45} to integrate the vehicle dynamics.

Referring to Fig. \ref{fig:merging}, CAVs arrive according to Poisson
processes with rates 2000 CAVs per hour and 1200 CAVs per hour for the main
and merging roads, respectively. The initial speed $v_{i}(t_{i}^{0})$ is also
randomly generated with a uniform distribution over $[15m/s,20m/s]$ at the
origins $O$ and $O^{\prime}$, respectively. The parameters for (\ref{eqn:ocbf}%
) and (\ref{VehicleDynamics}) are: $L_{2}=400m,L_{3}=407m,L_{4}%
=406.0622m,l=0.9378m,\varphi=1.8s,\delta=0m,u_{max}=3.924m/s^{2}%
,u_{min}=-5.886m/s^{2},v_{max}=30m/s,v_{min}=0m/s,\beta=1,\epsilon
=10,\Delta=0.1s,c=1,\Phi_{p}(x_{i}(t))=(\varphi+\frac{\delta}{v_{i}^{0}}%
)\frac{x_{i}(t)}{L_{p}}-\frac{\delta}{v_{i}^{0}},p\in\{1,2,3,4\}$. {We
consider all class $\mathcal{K}$ functions as cubic functions in
(\ref{eqn:cbf})} and { consider uniformly distributed noise processes (in [-2,
2] $m/s$ for $w_{i,1}(t)$ and in [-0.05, 0.05] $m/s^{2}$ for $w_{i,2}(t)$) for
all simulations}.

\begin{table}[ptb]
\caption{Comparison of OC, CBF and OCBF (with noise)}%
\vspace{-7mm}
\label{table:comp_OCBF_OC}
%	\label{table_example}
\par
\begin{center}%
\begin{tabular}
[c]{|c||c|c|c|c|c|}\hline
Method & $\alpha$ & Noise & Ave. time($s$) & Ave. $\frac{1}{2}u_{i}^{2}(t)$ &
Ave. obj.\\\hline\hline
CBF & N/A & no & 14.7539 & 19.7241 & N/A\\\hline\hline
Vissim & \multirow{3}*{$0.01$} & N/A & 31.5351 & 17.0415 &
19.2993\\\cline{1-1}\cline{3-6}%
\multirow{2}*{OCBF} &  & no & {\color{red} $\bm {22.6763}$} & {\color{red}
$\bm {6.7674}$} & {\color{red} $\bm {8.4458}$}\\\cline{3-6}
&  & yes & 22.7636 & 8.8133 & 10.4780\\\hline\hline
Vissim & \multirow{3}*{$0.25$} & N/A & 31.5351 & 17.0415 &
73.4767\\\cline{1-1}\cline{3-6}%
\multirow{2}*{OCBF} &  & no & {\color{red} $\bm {16.1588}$} & {\color{red}
$\bm {9.6914}$} & {\color{red} $\bm {38.3694}$}\\\cline{3-6}
&  & yes & 16.1811 & 11.2944 & 39.6146\\\hline\hline
Vissim & \multirow{3}*{$0.40$} & N/A & 31.5351 & 17.0415 &
107.3404\\\cline{1-1}\cline{3-6}%
\multirow{2}*{OCBF} &  & no & {\color{red} $\bm {14.4820}$} & {\color{red}
$\bm {14.6545}$} & {\color{red} $\bm {53.3915}$}\\\cline{3-6}
&  & yes & 14.4996 & 16.4412 & 54.5177\\\hline
\end{tabular}
\end{center}
\par
%\qquad  \quad $*$ m is a compulsory conversion from m/s.  	
\end{table}

We compare the simulation results between Vissim (human driver), the CBF
method \cite{Wei2019} (by setting $u_{ref}(t)=0$ and
$v_{ref}(t)=v_{max}$ in (\ref{eqn:ocbf})) and the OCBF method, as shown in
Table \ref{table:comp_OCBF_OC}. The CBF method is aggressive in travel time,
and thus has larger energy consumption than both the OCBF method and human
drivers. The OCBF method does better in both metrics than human drivers in
Vissim, and achieves about 50\% improvement in the objective function
(\ref{eqn:energyobj}) under all three different trade-off parameters $\alpha$
(recall that $\alpha$ trades off  travel time and energy in
(\ref{eqn:energyobja})).

\begin{figure}[ptbh]
	\vspace{-4.2mm}
\centering
\includegraphics[scale=0.4]{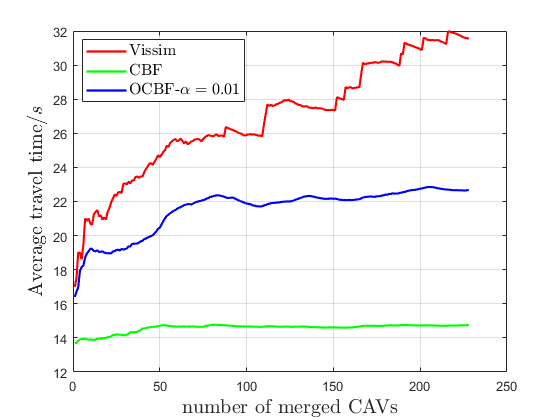} 
\vspace{-3mm}
\caption{Comparison of average travel time between Vissim, CBF and OCBF. }%
\label{fig:time}%
\end{figure}

\begin{figure}[ptbh]
	\vspace{-10mm}
\centering
\includegraphics[scale=0.4]{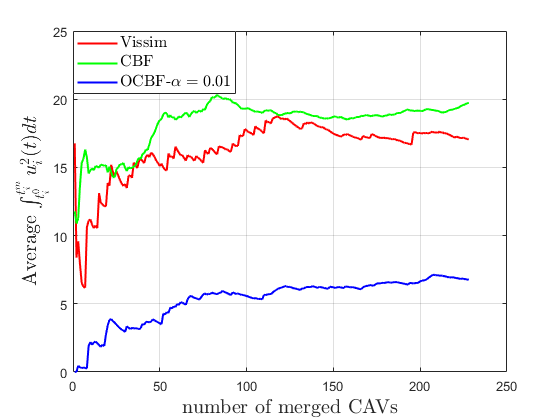} 
\vspace{-3mm}
\caption{Comparison of average energy
consumption profiles between Vissim, CBF and OCBF. }%
\label{fig:energy}%
\vspace{-3mm}
\end{figure}

In order to show whether the metrics have reached steady state or not, we
present the history of average travel time and energy consumption in Figs.
\ref{fig:time} and \ref{fig:energy}. The travel time in Vissim is still
increasing, indicating that traffic congestion is becoming worse. However,
both metrics in the CBF and OCBF methods are at steady state, providing
evidence of their ability to better manage traffic congestion.

\section{CONCLUSIONS}

\label{sec:conclude}

We have shown how to transform a multi-lane merging problem into a
decentralized optimal control problem, and combine OC with the CBF method to
solve the merging problem for CAVs in order to deal with cases where the OC
solution becomes difficult to obtain, as well as to handle the presence of
noise in the vehicle dynamics by exploiting the ability of CBFs to add
robustness to an OC controller. In addition, when considering more complex
objective functions for which analytical optimal control solutions are
unavailable, we can still adapt the CBF method to such objectives. Remaining
challenges include research on resequencing and extensions to large traffic
networks.
%\addtolength{\textheight}{-12cm}
%This command serves to balance the column lengths
%on the last page of the document manually. It shortens
%the textheight of the last page by a suitable amount.
%This command does not take effect until the next page
%so it should come on the page before the last. Make
%sure that you do not shorten the textheight too much.

%%%%%%%%%%%%%%%%%%%%%%%%%%%%%%%%%%%%%%%%%%%%%%%%%%%%%%%%%%%%%%%%%%%%%%%%%%%%%%%%

%%%%%%%%%%%%%%%%%%%%%%%%%%%%%%%%%%%%%%%%%%%%%%%%%%%%%%%%%%%%%%%%%%%%%%%%%%%%%%%%

%%%%%%%%%%%%%%%%%%%%%%%%%%%%%%%%%%%%%%%%%%%%%%%%%%%%%%%%%%%%%%%%%%%%%%%%%%%%%%%%

%%%%%%%%%%%%%%%%%%%%%%%%%%%%%%%%%%%%%%%%%%%%%%%%%%%%%%%%%%%%%%%%%%%%%%%%%%%%%%%%

\bibliographystyle{IEEEtran}
\bibliography{Hamilton}

\end{document}